\documentstyle[prb,epsfig,aps]{revtex}
\input epsf

\newcommand{\centps}[2]{
        \begin{center}
                \epsfig{file=#1,height=#2mm}
        \end{center}
}

\begin{document}
\draft

\twocolumn[\hsize\textwidth\columnwidth\hsize\csname @twocolumnfalse\endcsname

\title{Is the `Finite Bias Anomaly' in planar GaAs-Superconductor
        junctons caused by point-contact like structures?}

\author{S. Chaudhuri$^1$, P.F. Bagwell, D. McInturff, J.C.P. Chang, \\
S. Paak$^2$, M.R. Melloch and J.M. Woodall$^3$}
\address{School of Electrical and Computer Engineering}
\address{Purdue University, West Lafayette, Indiana 47907}
\author{T.M. Pekarek$^4$}
\address{Department of Physics}
\address{Purdue University, West Lafayette, Indiana 47907}
\author{B. C. Crooker}
\address{Department of Physics}
\address{Fordham University, Bronx, New York 10458}
\date{\today}
\maketitle

\begin{abstract}

We correlate transmission electron microscope (TEM) pictures of
superconducting In contacts to an AlGaAs/GaAs heterojunction with
differential conductance spectroscopy performed on the same
heterojunction. Metals deposited onto a (100) AlGaAs/GaAs
heterostructure do not form planar contacts but, during thermal
annealing, grow down into the heterostructure along crystallographic
planes in pyramid-like `point contacts'.  Random surface nucleation
and growth gives rise to a different interface transmission for each
superconducting point contact.  Samples annealed for different times,
and therefore having different contact geometry, show variations in
$dI/dV$ characteristic of ballistic transport of Cooper pairs, wave
interference between different point emitters, and different types of
weak localization corrections to Giaever tunneling.  We give a
possible mechanism whereby the `finite bias anomaly' of Poirier et
al. (Phys. Rev. Lett., {\bf 79}, 2105 (1997)), also observed in these
samples, can arise by adding the conductance of independent
superconducting point emitters in parallel.

\end{abstract}

\pacs{PACS 74.80Fp, 73.20.Dx, 74.50+r}

]  \narrowtext
\flushbottom

\section{Introduction}
\indent

The current-voltage (I-V) relation of normal metal -superconductor
(NS) interfaces~\cite{btk}-\cite{lesovik} is strongly modified both by
wave interference phenomena (producing quasi-bound Andreev levels) and
ballistic transport at the interface. Several groups have observed
such novel superconducting phenomena at the NS interface between a
superconductor and semiconductor~\cite{kleins1}-\cite{poirier}.  Based
on the success of Marsh et al.~\cite{marsh1}-\cite{marsh4} in
fabricating In and Sn `alloyed' contacts to a two-dimensional electron
gas (2DEG) formed at the AlGaAs/GaAs interface, we have studied
`alloyed' In contacts to the 2DEG.  We find the mechanism for
producing highly transmissive NS contacts is In growth into the AlGaAs
`guided' along a preferred crystallographic direction. For an
AlGaAs/GaAs heterojunction with a [100] oriented surface, we find that
In growth into the AlGaAs occurs preferentially along the \{111\}
crystallographic planes. This unusual type of In growth into the
AlGaAs produces an `inverted pyramid' or `field emission' point
contact tip as shown in Fig.~\ref{cartoon1}(a). Similar microstructure
for metallic contacts to GaAs has been observed for both
AuGeNi~\cite{triangles} and Au~\cite{gold-GaAs} metallizations.  This
guiding of In into the AlGaAs also allows the In to maintain its
superconducting properties. An AlInGaAs alloy, formed by diffusing In
into AlGaAs, would simply be a normal metal.

For such a crystallographically defined point contact metallization,
we find the closer one can grow the tip of the point contact to the
2DEG without contacting it, the higher the transmission coefficient of
an electron incident from the NS contact into the 2DEG. For such
nearly ballistic transport through the NS junction, a corresponding
excess current results~\cite{btk}. Growing the In down into direct
contact with the 2DEG, on the other hand, results in a
low-transmission normal metal - insulator - superconductor (NIS)
contact and its corresponding Giaever tunneling I-V
characteristic~\cite{btk}. We postulate that In in direct contact with
the 2DEG depletes the electrons around it, forcing the superconducting
electrons to tunnel through a large depletion layer near the contact,
shown schematically in Fig.~\ref{cartoon1}(b). Electron depletion
around a metallic contact to GaAs is commonly known as a Schottky
barrier.  A similar mechanism for forming highly transmissive AuGeNi
contacts to GaAs was originally postulated by
Braslau~\cite{braslau}-\cite{woodall}. This mechanism for
forming highly transmissive tunneling type AuGeNi contacts to GaAs is
the reason some AlGaAs/GaAs heterojunction transistors can operate at
low temperatures with low contact resistance.

\begin{figure}
\centps{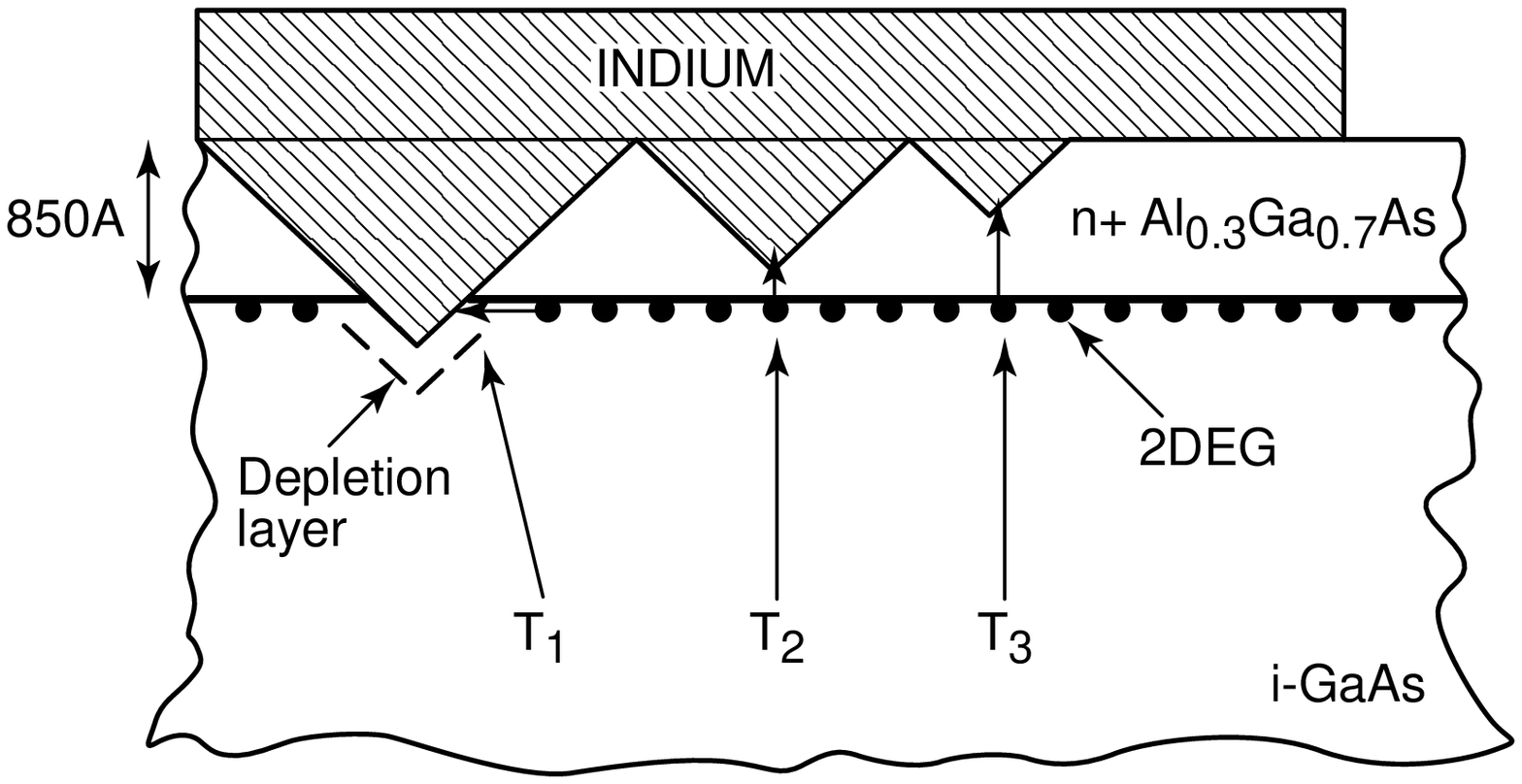}{45}
\centps{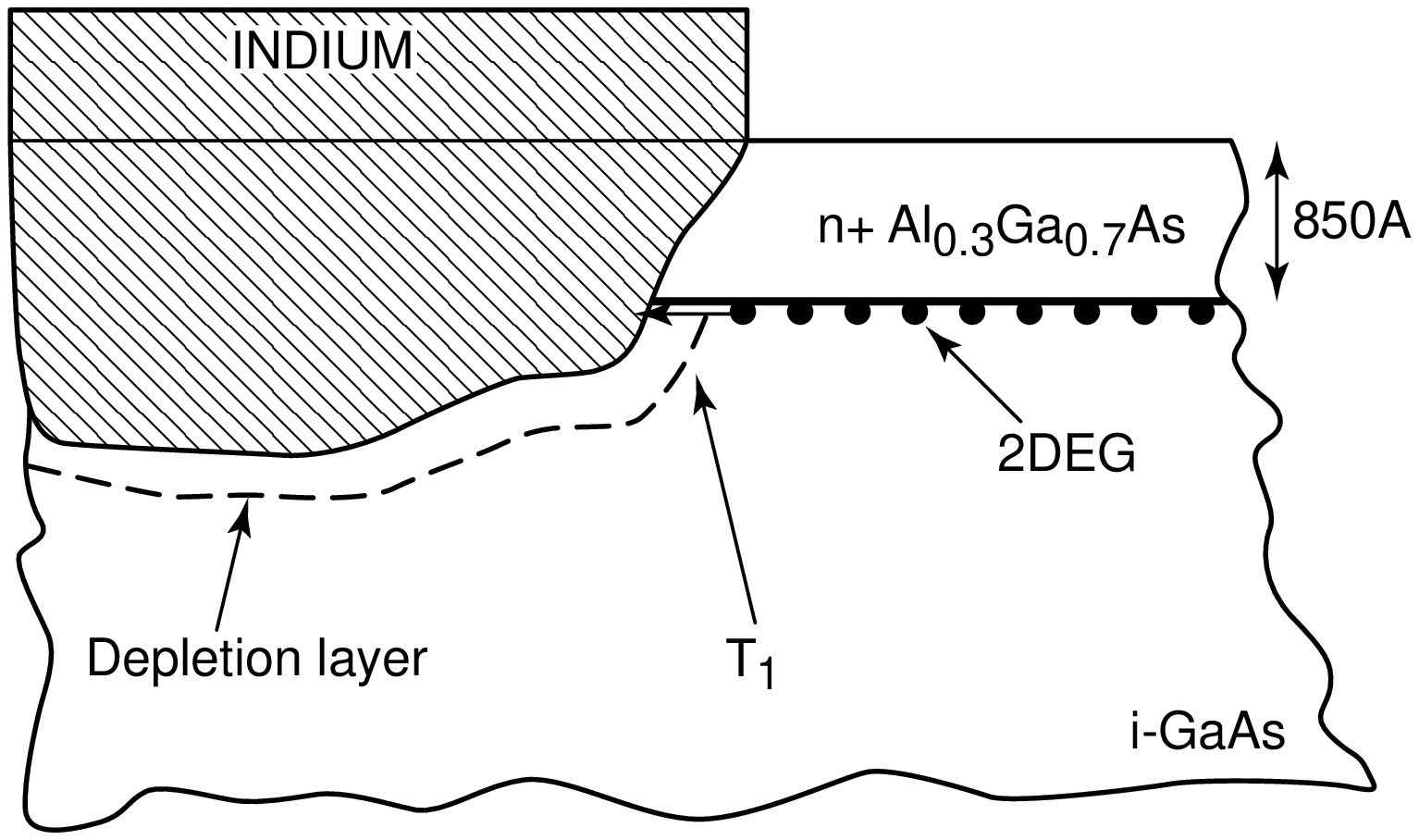}{45}
\caption{Schematic of the In profile obtained after annealing for (a)
a low temperature/short time anneal (Sample 1) and (b) a higher
temperature/longer time anneal (Sample 2). Penetration of In into
the AlGaAs layer is guided along a preferred crystallographic direction
in (a), forming point contacts to the 2DEG. The point contacts grow 
together and penetrate the 2DEG in (b).}
\label{cartoon1}
\end{figure}

In this paper we correlate transmission electron microscope (TEM)
photographs of the superconducting In contacts to the resulting I-V
characteristics of the NS junctions. All the different $dI/dV$
characteristics shown in this paper are from nominally identical
samples, grown and prepared from the same GaAs wafer at the same
time. The only differences between the samples is in post process
contact annealing, and hence in the contact geometry.

Changes in contact geometry produce widely different $dI/dV$
characteristics.  In point contacts grown near the 2DEG result in
ballistic transport of electrons through the NS contact and an excess
current.  In in direct contact with the 2DEG produces lower
transmission contacts and Giaever tunneling. Since the GaAs
semiconductor forming the normal metal is also weakly localized, we
are able to observe weak localization corrections to Giaever
tunneling~\cite{vanwees1}.  In In/GaAs junctions where some region of
the contact is transmissive, that portion of the contact will produce
a conductance drop around zero bias~\cite{marmorkos1}.  The incoherent
addition of the conductance from different regions of the same
`contact' could therefore generate the `finite bias anomaly' seen in
Ref.~\cite{poirier}.

\section{Formation of the NS Contact}
\indent

A cross section of the unnannealed In/GaAs heterostructure is shown in
Fig.~\ref{heterostructure}(a). An undoped Al$_{0.3}$Ga$_{0.7}$As
spacer layer, followed by a Si doped Al$_{0.3}$Ga$_{0.7}$As layer and
a 50 $A^{o}$ protective Si doped GaAs layer, was grown on an undoped
(semi-insulating) (100) GaAs substrate. The resulting mobility of the
2DEG at liquid nitrogen temperature was about 125,000 $cm^{2}$/$V-s$.
The In contacts were deposited by thermal evaporation and liftoff.  A
top view of the device is shown in Fig.~\ref{heterostructure}(b).  Two
In pads, each of dimension 3.2mm$X$2.5mm and seperated by a nominal
gap of 4$\mu$, were deposited on top of the heterostructure using
thermal evaporation lift off. After annealing, we diffused In
through the AlGaAs barrier layer to contact the 2DEG.
The annealing temperature was varied between 500$^{o}$C and
600$^{o}$C.  For temperatures of 450$^{o}$C or less the In failed
to contact the 2DEG, while annealing temperatures greater than
700$^{o}$C caused thermal deterioration of the interface.  Large In
grains, 1-2 microns in diameter, grew on the contacts after annealing.

\begin{figure}
\centps{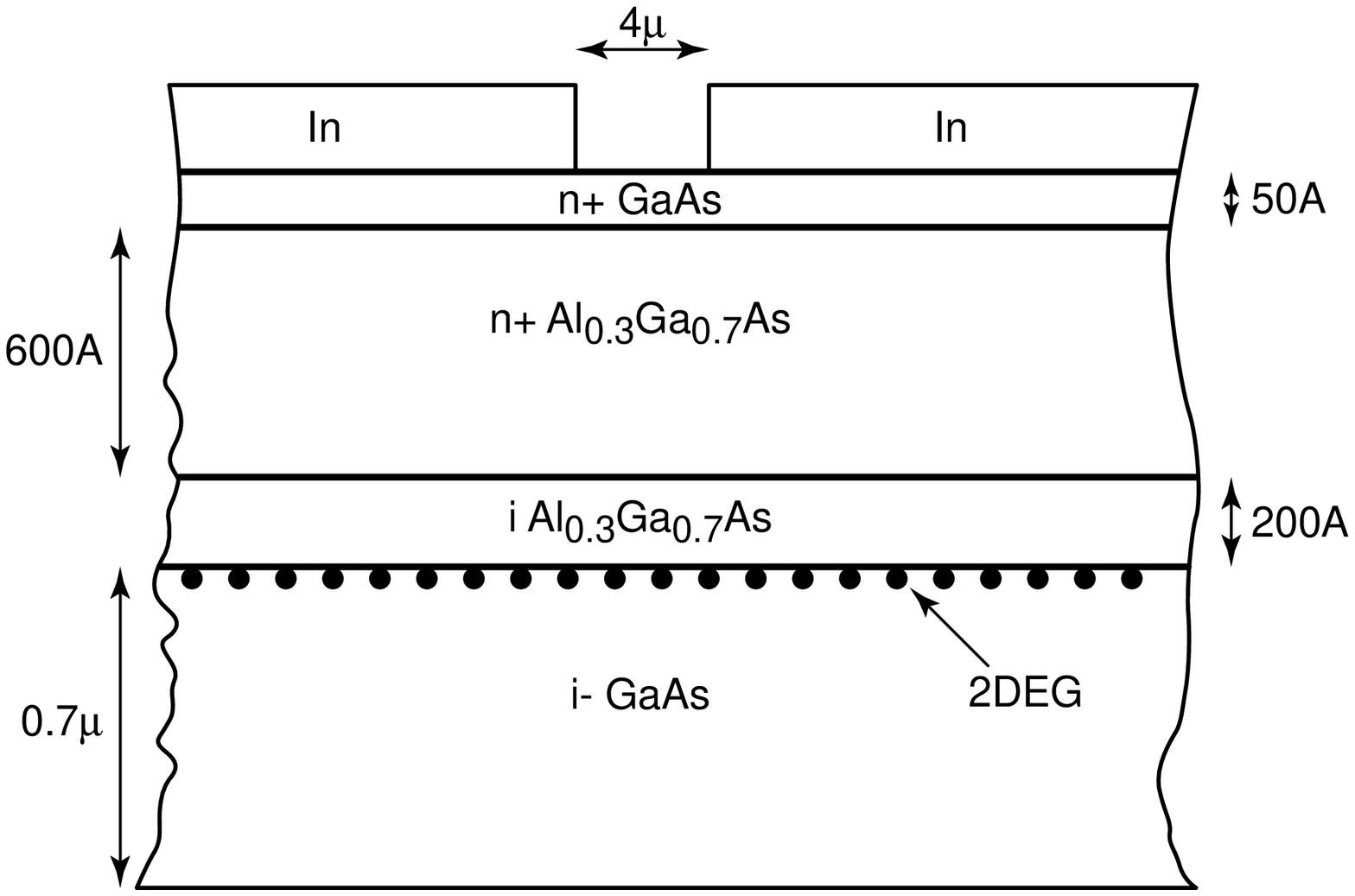}{45}
\centps{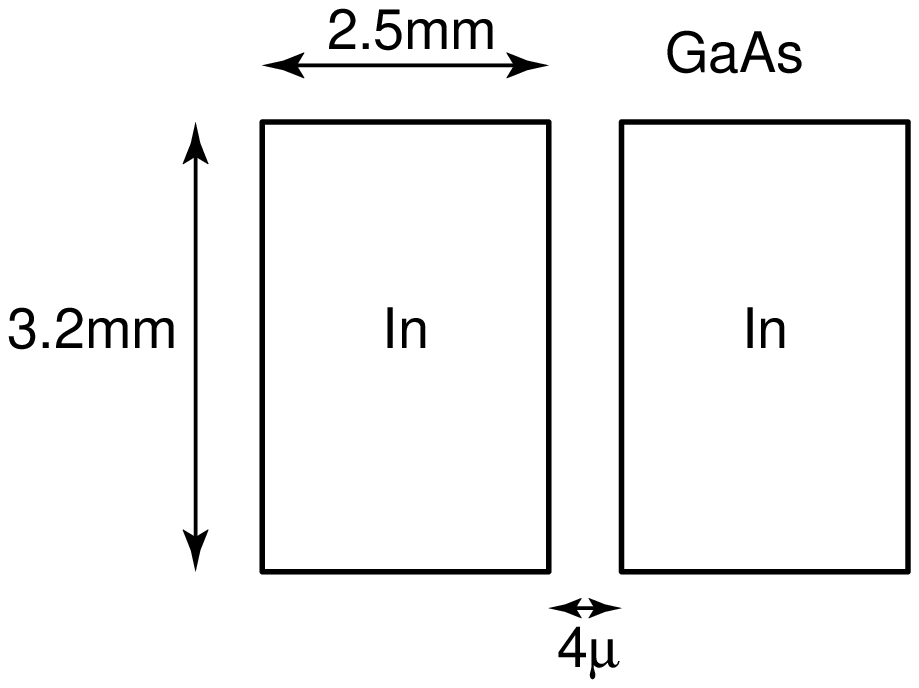}{45}
\caption{(a) Cross-section of the AlGaAs/GaAs heterostructure before
annealing the superconducting In contacts. (b) Top view of the
contact geometry.}
\label{heterostructure}
\end{figure}

A TEM micrograph of Sample 1, annealed for a relatively short duration
e.g. at 550 $^{o}C$ for 2 minutes, is shown in Fig.~\ref{tem1}(a).
Fig.~\ref{tem1}(a) shows that Indium starts growing into (100)
AlGaAs/GaAs preferentially along the $<$111$>$ directions. The growth
seems to be guided by the \{111\} crystallographic planes of GaAs
inclined at angles of about 55$^{o}$ to the surface. Therefore for
short annealing times we get `spikes' of Indium descending towards the
interface, forming an array of point contacts pictured in
Fig.~\ref{tem1}(a). The spikes are rather non-uniform in size and
irregular in their penetration depths, probably nucleating at defects
in the suface oxide of AlGaAs.  In grain growth on the wafer surface
therefore has little effect on the final microstructure and geometry
of the NS contact.  The net total conductance of the junction is
determined by the sum of the conductance of all the point contacts in
parallel.  This growth mechanism is similar to that for alloyed
Au-Ge-Ni contacts with GaAs~\cite{woodall} in which most of the
conduction is through isolated Ge rich islands formed on the GaAs.

\begin{figure}
	\begin{center}
                \setlength{\epsfxsize}{7.0truecm}
                \setlength{\epsfysize}{7.0truecm}
                \epsfbox{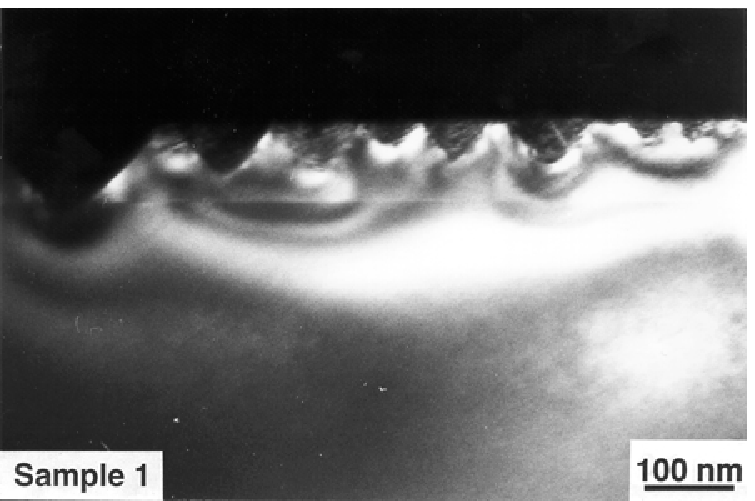}
                \setlength{\epsfxsize}{7.0truecm}
                \setlength{\epsfysize}{7.0truecm}
                \epsfbox{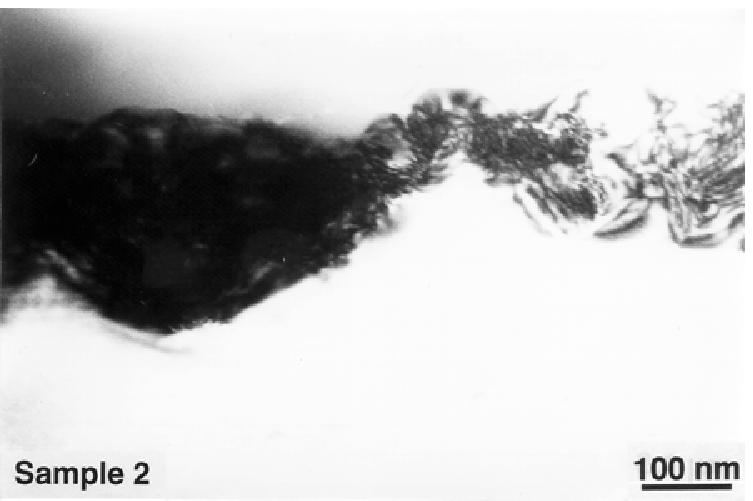}
	\end{center}
\caption{TEM photographs of an In-AlGaAs/GaAs contact annealed (a) for
a short time (Sample 1) and (b) a longer time (Sample 2).  Penetration
of In into the sample is guided by the \{111\} planes, forming the In
point contacts are clearly observable in (a). The point contacts
agglomerate and grow through the AlGaAs/GaAs interface in (b).}
\label{tem1}
\end{figure}

A TEM micrograph of Sample 2, annealed at 550$^{o}$C for 6 minutes and
then at 650$^{o}$C for 3 minutes, is shown in Fig.~\ref{tem1}(b). For
the longer annealing times and higher annealing temperatures used in
sample 2, more such In spikes grow from the deposited In
contact. Fig.~\ref{tem1}(b) shows these In spikes coalesce and
penetrate completely through the AlGaAs/GaAs interface to physically
touch the 2DEG, as depicted schematically in Fig.~\ref{cartoon1}(b).
The conductance characteristics of Samples 1 and 2 differ dramatically
as described in the next section.

A high magnification TEM photograph of one of the inverted pyramid
type In spikes in Sample 1 is shown in Fig.~\ref{tem2}. The (100) GaAs
surface is towards the top of Fig.~\ref{tem2}. The penetration of In
into the AlGaAs is clearly guided by \{111\} crystallographic
planes. One can see in the high resolution TEM picture that the In
indeed follows the \{111\} AlGaAs planes for several atoms. The
In-AlGaAs boundary then moves abruptly along the [010] direction for
1-2 atoms before continuing along the \{111\} planes. The detailed TEM
picture in Fig.~\ref{tem2} shows that while the \{111\} planes
strongly guide the growth of In into (100), the In does not exactly
follow those planes.  However, the overall structure of the In
contacts still resembles a point contact.

\begin{figure}
	\begin{center}
                \setlength{\epsfxsize}{7.0truecm}
                \setlength{\epsfysize}{7.0truecm}
                \epsfbox{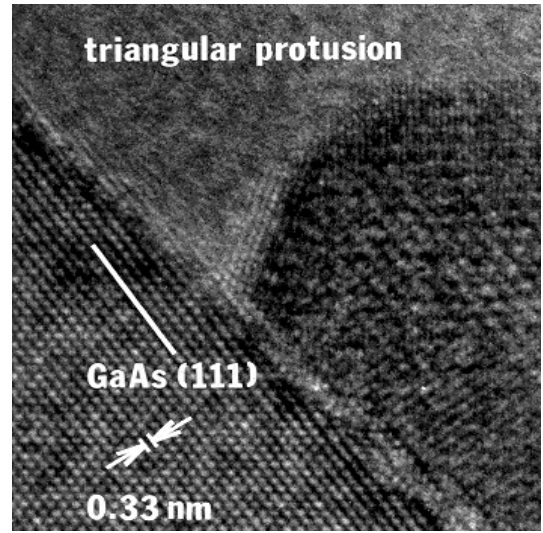}
	\end{center}
\caption{High magnification TEM photograph of a portion of a single In
point contact grown into a (100) AlGaAs surface.  In growth into the
AlGaAs is strongly guided along the \{111\} planes.}
\label{tem2}
\end{figure}

\section{Current-Voltage Characteristics}
\indent

Hall measurements were made on the 2DEG at the interface at a
temperature of 400 mK. From the slope of transverse resistance
$R_{xy}$ we estimate the carrier density to be $1.8X10^{11}$
$cm^{-2}$. The mobility was 215,000 ${{cm}^2}/{V-s}$ which yielded a
mean free path $l$ of about 0.4 $\mu$.  From the low-field
magnetoresistance, we estimate the phase breaking length $l_{\phi}$ to
be about 1.6 $\mu$m~\cite{chaud-thesis}, which is less than the 4
$\mu$m separation between the two In pads in
Fig.~\ref{heterostructure}(b). We conclude that the conductance of the
devices is essentially equivalent to two NS junctions in series
separated by a series resistor (the 2DEG).

\subsection{NS Junction with an excess current}
\indent

In Fig.~\ref{base12}(a) we show the conductance characteristics of
Sample 1 at a base temperature of 100 mK.  Fig.~\ref{base12}(a) shows
an increase in conductance of about 10-12\% around a range of $\pm$
6mV.  The 10-12\% excess conductance around zero voltage shows that
the junction behaves like a moderately transmissive
junction~\cite{btk}, with the majority of the diffused Indium spikes
forming transmissive interfaces with the 2DEG of the GaAs.  The large
oscillations of the conductance in Fig.~\ref{base12}(a) are
reproducible with thermal recycling of the device.  The oscillations
are therefore likely a consequence of electron wave interference due
to scattering from the fixed In point contacts to the 2DEG.  The
parasitic resistance of the 2DEG in series with the two NS junctions
stretches the dI/dV versus V characteristics along the voltage axis,
and also suppresses any changes observed along the dI/dV axis.  Series
resistance of the 2DEG explains why we observe only a 10-12\% excess
conductance instead of an excess conductance approaching 100\% in
Fig.~\ref{base12}(a).

\begin{figure}
\centps{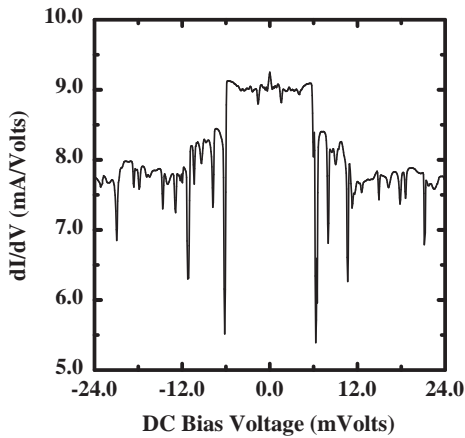}{60}
\centps{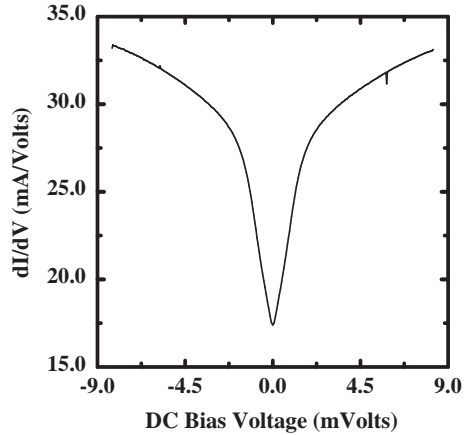}{60}
\caption{Conductance vs the DC bias voltage across the two Indium pads
for (a) Sample 1 and (b) Sample 2. Sample 1 shows nearly 100\%
ballistic transport of electrons (after correcting for series
resistance), while Sample 2 shows Giaever tunneling (through a 
tunnel barrier having transmission of the order $T \simeq 0.1$).}
\label{base12}
\end{figure}

We can crudely estimate the interface transmission in
Fig.~\ref{base12}(a) by comparing it to an ideal, ballistic NS
junction~\cite{btk}. A perfectly transmissive interface shows a 100\%
rise of conductance when the voltage bias satisfies $|eV| \leq
\Delta$, where $2\Delta$ superconducting energy
gap~\cite{btk}. Setting $2 \Delta =$ 6mV (for two NS junctions in
series), and using the BCS formula $2 \Delta = 3.5 k_B T_c$, gives a
critical temperature $T_c = 20$K. Since the actual critical
temperature of the contacts is about $3.4$K, we infer a 1mV drop
across the two NS interfaces and 5mV across the semiconductor.  The
total conductance at the gap voltage can be read directly from
Fig.~\ref{base12}(a) as about 9mS, from which we infer a series
resistance of about 92.6$\Omega$ and a resistance of the two NS
interfaces in series of about 18.5$\Omega$ (when the voltage across
the NS interface is less than the energy gap). When the voltage across
the two NS interfaces is large, we can read directly off
Fig.~\ref{base12}(a) a conductance of about 7.75mS, or a total
resistance of about 129$\Omega$. Subtracting the series resistance, we
see the device resistance changes from 18.5$\Omega$ when $V_{\rm
interface} \leq \Delta$ to about 36.4$\Omega$ when $V_{\rm interface}
\gg \Delta$, roughly a 97\% increase in background conductance.  Since
strong wave interference is present in Fig.~\ref{base12}(a), this
comparison should be regarded as giving an order of magnitude
estimate. However, after correcting for series resistance, this
estimate indicates the junction in Fig.~\ref{base12}(a) is a nearly
ballistic NS interface.

\subsection{Giaever-type Tunnel Junctions}
\indent

In Fig.~\ref{base12}(b) we show the differential conductance for
Sample 2 at a temperature of 400mK. The differential conductance is
suppressed around zero voltages, indicating Giaever tunneling and a
lower transmission contact. We can conclude from Figs.~\ref{cartoon1}
and \ref{base12} that In in intimate contact with the 2DEG forms a
relatively low transmission contact, whereas In nearby but not
directly in contact with the 2DEG forms a high transmission contact.
The temperature dependence of the differential conductance for Samples
1 and 2 is also consistent with a ballistic contact and a low
transmission contact, respectively. The differential conductance of
Sample 1 is relatively constant with temperature, while the
differential conductance of Sample 2 greatly decreases as the
temperature is lowered (indicating thermionic emission).

We have taken several NS contacts displaying Giaever tunneling at low
temperature and annealed them for longer times, attempting to obtain a
ballistic NS interface. In all cases further annealing simply makes
the Giaever tunneling characteristic more pronounced, indicating that
further annealing lowers the interface transmission. The room
temperature conductance of the sample improves with further annealing,
however, simply due to an increase of the effective contact
area. Since the room temperature conduction mechanism through the
contact of Sample 2 is thermionic emission, it simply scales with the
contact area. Further annealing improves the room temperature
conductance, but worsens the low temperature conductance. There is an
optimal annealing time where In grows down to almost reach the 2DEG,
but is not in physical contact with the 2DEG. Any further annealing
after this point degrades interface transmission.

\subsection{Effect of Weak Localization on Giaever Tunneling}
\indent

For an NS junction of length L, obeying the condition $l\ll L \leq
l_{\phi}$, electrons which initially failed to Andreev refelct from
the NS interface can backscatter again to the NS interface.
Therefore, weak localization inside the normal conductor gives the
electrons additional opportunities for Andreev reflection.  The net
effect of Giaever tunneling at the NS junction combined with
weak localization inside the normal conductor is an enhancement of the
total Andreev reflection probability at the Fermi level, leading to an
additional conductance peak around zero bias
voltage~\cite{kleins1,vanwees1}.

Fig.~\ref{weakloc} shows the differential conductance for the NS
junction annealed at 550$^{o}$C for 3 minutes, which we call Sample
3. Sample 3 displays the conductance peak around zero bias voltage
first observed in Ref.~~\cite{kleins1} and explained in
Ref.~\cite{vanwees1}. Disorder assisted backscattering can cause a
zero bias conductance peak of magnitude up to 10\% of the background
conductance value. In our case the zero bias peak is about a 1.5\%
increase over the background conductance. The Giaever tunneling
feature is also spread over a large voltage range larger than 1mV.
Parasitic resistance from the 2DEG again explains the smaller zero
bias conductance peak and the spreading out of the dI/dV versus V
along the voltage axis.  Since the height of the zero bias conductance
peak saturates by 300 mK Fig.~\ref{weakloc}(a), the peak is not the
precursor of a supercurrent between the two contacts.  The
supercurrent should be negligibly small in any case, since the sample
satisfies $l\ll L \geq l_{\phi}$. We adequately filtered RF noise away
from the devices to observe supercurrents in other superconductor -
semiconductor samples with closer pad separation.  We would also have
observed such a supercurrent if it were present in this sample.

Fig.~\ref{weakloc}(b) shows the magnetic field dependence of the
conductance for Sample 3. For a junction of length L and width W,
where $L,W > L_{\phi}$, the magnetic field required to destroy this
zero bias conductance peak is of the order
$B_{c}={\phi_{0}}/{{L_{\phi}}^2}$. In Sample 3 we observe $B_{c}
\simeq$ 80 Gauss. The calculated field is $B_{c} \simeq$ 20 Gauss,
nearly four times smaller than the observed value. It is possible that
the $L_{\phi}$ is overestimated, i.e. it may be around
$L_{\phi}\simeq$ 0.8 $\mu$m. Since these numbers are not precise data
fits based on any quantitative theory, it is comforting that we obtain
roughly the coherence length obtained by previous weak localization
measurements on the 2DEG.  The shift in the background conductance
with the magnetic field in Fig.~\ref{weakloc}(b) is also due to the
parasitic magnetoresistance of the 2DEG.

\begin{figure}
\centps{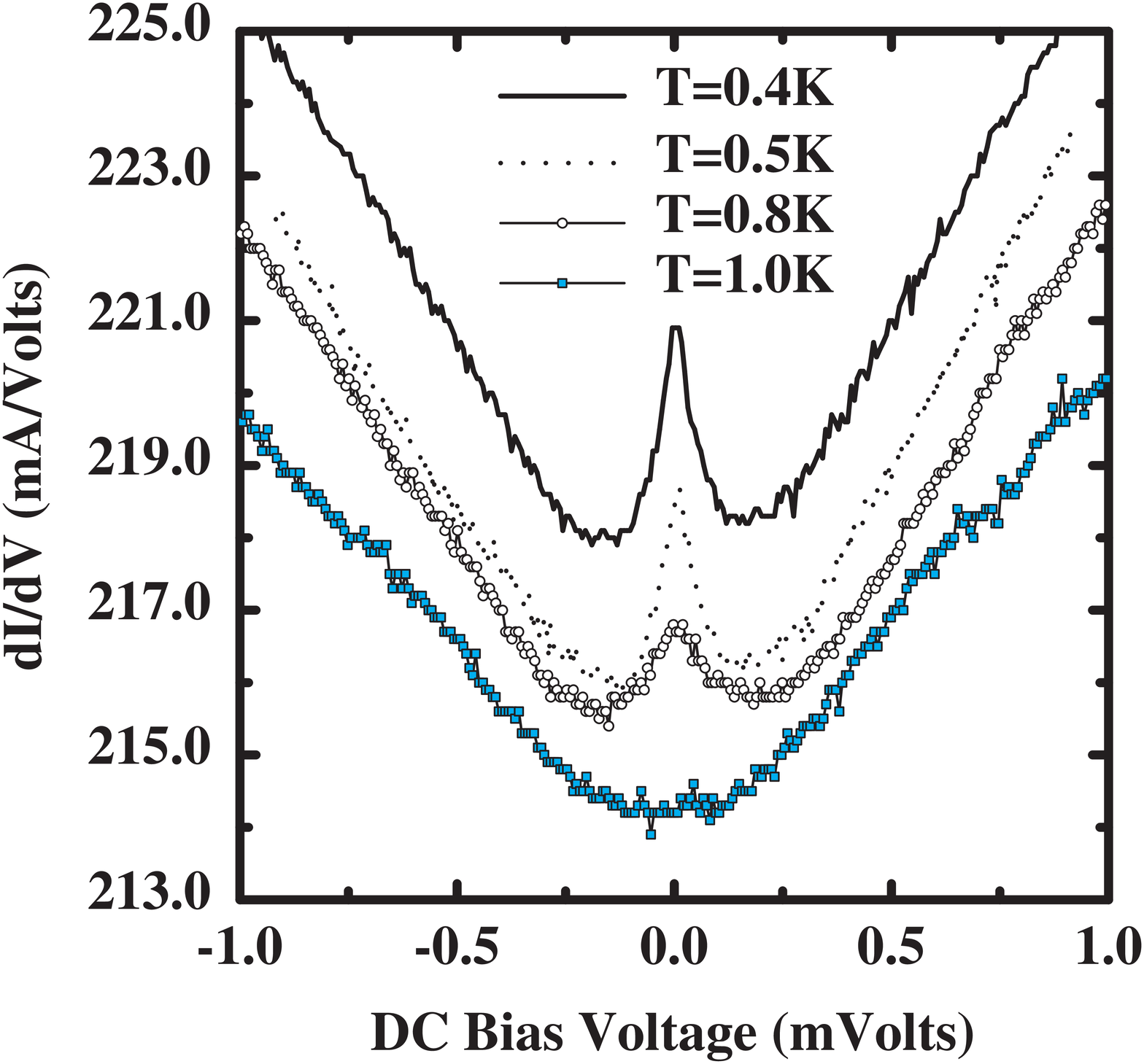}{60}
\centps{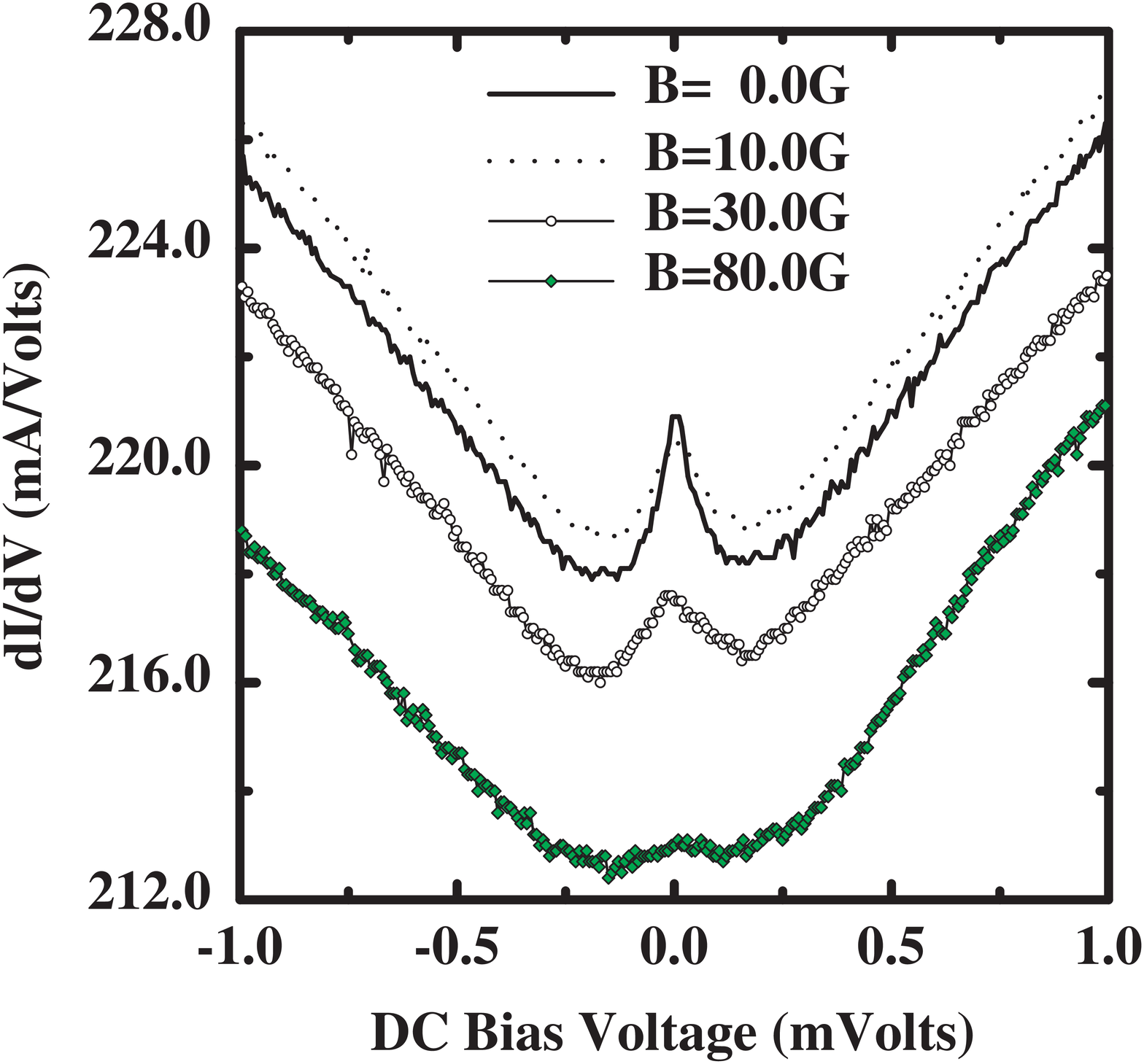}{60}
\caption{ Conductance vs the DC bias voltage for Sample 3 for (a)
different temperatures and (b) different magnetic fields. A
conductance peak develops near zero bias voltage, a correction to
Giaever tunneling arising from weak localization inside the normal
metal.}
\label{weakloc}
\end{figure}

\subsection{Anamalous Weak Localization Corrections to Giaever
Tunneling}
\indent

On most samples where we observed weak localization corrections to
Giaever tunneling, we obtained conductance characteristics similar to
those in Fig.~\ref{weakloc}. However, Fig.~\ref{anomloc} shows the
conductance characteristics for an NS junction annealed at 500$^{o}$ C
for 2 minutes which we call Sample 4.  Overall Sample 4 displays a
background of Giaever tunneling. A zero bias conductance peak (similar
to the one in Fig.~\ref{weakloc}) continues to develop for
temperatures down to about 800 mK in Fig.~\ref{anomloc}(a).  At a
temperature of 650 mK in Fig.~\ref{anomloc}(a), however, a conductance
dip begins developing around zero bias. This dip in conductance around
zero bias, which is superposed on the broader conductance peak, is
nearly fully developed by 300 mK as shown in Fig.~\ref{anomloc}(a).
There is little change in the differential conductance between 300 mK
and 180 mK in Fig.~\ref{anomloc}(a). This anomalous dip feature
superimposed on the weak localization correction to Giaever tunneling
is reproducible on thermally cycling the NS junction back to room
temperature and again down to mK temperatures.

\begin{figure}
\centps{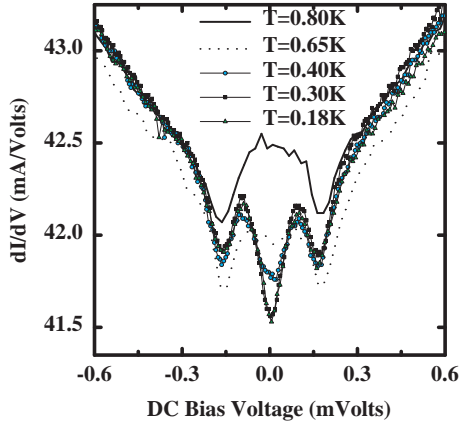}{60}
\centps{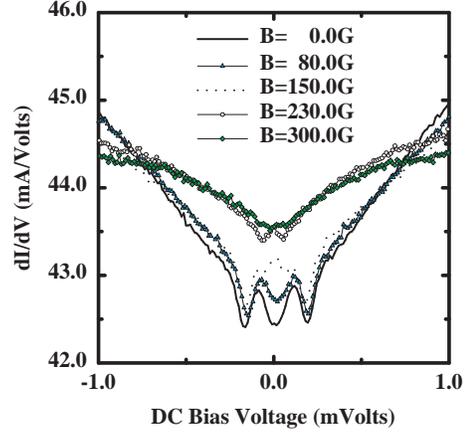}{60}
\caption{ Conductance vs the DC bias voltage for Sample 4 for (a)
different temperatures and (b) different magnetic fields. The additional
conductance dip which develops around zero bias voltage can be explained by an
inhomogeneous NS contact consisting of both high and low transmission regions. }
\label{anomloc}
\end{figure}

Marmorkos, Beenakker, and Jalabert~\cite{marmorkos1} have numerically
simulated the conductance of an NS junction in contact with a dirty
normal metal. For low transmission interfaces they numerically
observe, in Fig.~2 of Ref.~\cite{marmorkos1}, the zero bias
conductance peak associated with the weak localization corrections to
Giaever tunneling.  However, for high transmission between the NS
interface and normal conductor, the numerical simulations of
Ref.~\cite{marmorkos1} reveal that the conductance peak changes into a
conductance dip around zero bias.  Ref.~\cite{marmorkos1} therefore
shows that the same weak localization phenomena which causes a zero
bias conductance peak in low transmission contacts causes a zero bias
conductance dip for highly transmissive NS interfaces.

The numerical simulation in Ref.~\cite{marmorkos1} offers one possible
way to explain the conductance dip around zero bias we observe in
Sample 4.  The overall conductance of Sample 4 displays Giaever
tunneling. Therefore, the majority of the NS interface area in Sample
4 has an additional tunneling barrier between the superconductor and
2DEG, namely the depletion region shown in
Fig.~\ref{cartoon1}(b). However, the type of NS junctions we form
by diffusion In into AlGaAs/GaAs are inhomogeneous enough that a
significant fraction of the sample can form a transmissive NS
interface of the type shown in Fig.~\ref{cartoon1}(a). The
conductance we observe in Fig.~\ref{anomloc} will be a parallel
combination of these two different types of NS junctions, as shown
schematically in Fig.~\ref{cartoon2}. For voltages away from $V=0$,
the slow variation of the background Giaever tunneling conductance
dominates the dI/dV curve. For voltages very close to zero bias, the
weak localization phenomena at the transmissive regions dominate and
leads to the observed conductance peak and dip.

\begin{figure}
\centps{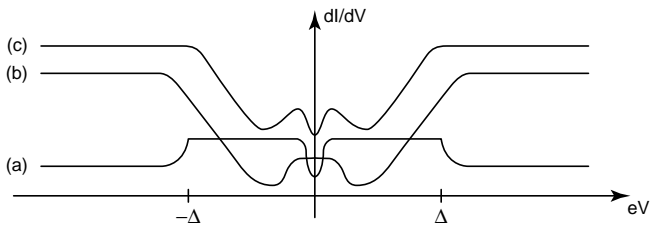}{30}
\setlength{\epsfysize}{7.0truecm}
\caption{Possible explanation for conductance dip around zero voltage
observed inside the zero bias conductance peak. Low transmission
regions of the interface give both the overall Giaever tunneling shape
of the dI/dV versus V and the zero bias conductance peak. A few high
transmission regions of the contact could produce the zero bias
conductance dip.}
\label{cartoon2}
\end{figure}

The peak at finite bias in Figs.~\ref{anomloc} was first observed by
Poirier et al.~\cite{poirier}, who called it the `finite bias
anomaly'. One problem with using the simulations of Mormorkos et
al.~\cite{} to explain a conductance dip around zero bias is that it
requires a high interface transmission, whereas the data of Poirier et
al.~\cite{poirier} (and our own data) show a Giaever tunneling
background (low average interface transmission). The inherent
inhomogeneity of supposedly planar superconducting In contacts to the
2DEG in AlGaAs/GaAs we have demonstrated in this paper overcomes this
difficulty. A few high transmission point emitters can produce the
conductance dip around zero bias, whereas the majority of the contact
can maintain low overall interface transmissivity. The weak localization
dip around zero bias can therefore peacefully coexist with a Giaever
tunneling background conductance.

The weak localization correction to the conductance of a ballistic NS
junction could have been more clearly observed in Sample 1, were it
actually present in that sample. Similarly Sample 2 (and several other
samples we measured) did not exhibit the weak localization correction
to the Giaever tunneling conductance. The exact impurity configuration
near a particular NS interface will determine whether or not the weak
localization correction to the conductance appears in any given
sample.  Perhaps it is therefore not surprising that the weak
localization correction to the conductance can be observed only in a
fraction of the samples.

A different mechanism which splits the zero bias conductance peak in
NI$_1$NI$_2$S junctions was developed in Ref.~\cite{lesovik}.  Weak
localization inside the middle N region produces the zero bias
conductance peak. If the two insulators I$_1$ and I$_2$ have two
different transmission coefficients, the zero bias conductance peak is
split as shown in Fig.~3 of Ref.~\cite{lesovik}. These two barriers,
having different transmissivity, is the same mechanism proposed by
Poirier et al.~\cite{poirier} to account for the `finite bias anomaly'.
Ref.~\cite{poirier} proposed a model which used the Schottky barrier at
the NS interface to produce I$_2$, and an impurity inside the
semiconductor as I$_1$.  The spacing between I$_1$ and I$_2$ is L, a
random number set by the impurity configuration. The McMillan-Rowell
resonance nearest the Fermi level survives in the conductance of an
NI$_1$NI$_2$S junction upon averaging over different L, producing a
finite bias anomaly whose voltage is set by the average L.

The composite `point emitter' model for the contact developed in this
paper may also provide some support for this NI$_1$NI$_2$S model for
the `finite bias anomaly'. An electron moving through the 2DEG past a
point emitter would see that emitter as a scattering center,
equivalent to an insulating barrier. The distance between the emitters
in Sample 1 is of the order 100nm, less than the electron phase
coherence length. In Sample 1, therefore, could be regarded as a type
of NI$_1$NI$_2$NI$_3$~...~S junction. Each normal metal region N would
also be weakly localized. This model may also produce a finite bias
anomaly, but would require further numerical support. A two-dimensional
numerical simulation, where the electrons could actually move around the
point emitter scattering centers, would be required to confirm this
picture. 

\section{Conclusions}
\indent

We have measured the differential conductance of superconductor-normal
metal junctions formed by diffusing Indium into AlGaAs/GaAs
heterostructures. In grows into a AlGaAs/GaAs heterostructure having a
[100] oriented surface preferentially along the \{111\}
crystallographic planes. Instead of a planar diffusion profile, we
therefore find that In forms `inverted pyramids' or point contacts to
the 2DEG. Supposedly `planar' superconducting In contacts to the
electron gas in an AlGaAs/GaAs heterojunction are therefore actually
composed of many point emitters.  Correlating the contact
microstructure observed on different samples with the differential
conductance spectroscopy of the NS contact allowed us both to explain
many observed features in the conductance and to determine the
mechanism of superconducting (ohmic) contacts to the 2DEG in this
materials system.

For NS junctions annealed at a moderate temperature for a short times,
so that the In point contacts do not physically touch the 2DEG, we
obtain highly transmissive NS junctions. Due to the contact
inhomogeneity, the point emitters nucleate and grow at different rates
into the semiconductor. We observed wave interference between these
different superconducting emitters in transmissive NS junctions.  For
identically prepared NS junctions annealed at higher temperatures and
for longer times, so that the In point contacts grow together and have
direct physical contact with the 2DEG, we obtain a lower transmission
NS interface and Giaever tunneling. This is due to a depletion layer
which forms around the In which directly touches the 2DEG.  Further
annealing simply increases the effective strength of the interface
barrier between N and S, as regions of the In which previously were
not in direct physical contact with the 2DEG come in contact with the
2DEG.

Since the semiconductor forming N is disordered and has a reasonable
phase coherence length, weak localization corrections to the
differential conductance around zero bias voltage are also observed in
this materials system. This zero bias conductance peak is a correction
to Giaever tunneling which has been previously observed by several
other groups~\cite{kleins1,vanwees1}. We observed an additional dip
inside this zero bias conductance peak which develops in some samples
at low temperature~\cite{poirier}. One possible explanation for the
additional dip is due to contact inhomogeneity, where a small
percentage of the contact is a nearly ballistic NS interface while
most of the NS contact area remains in the tunneling limit. This
conductance dip around zero bias voltage is therefore possible
evidence for the predicted weak localization correction to the
conductance of ballistic normal metal - superconductor junctions in
Ref.~\cite{marmorkos1}. In any case, explanations for this `finite
bias anomaly' should account for the actual non-planar physical
structure of the superconducting contact.

\section{Acknowledgments}
\indent

We gratefully acknowledge support from the David and Lucile Packard
Foundation and from the MRSEC of the National Science Foundation under
grant No. No. DMR-9400415.  We thank Tamer Rizk, Richard Riedel, Manoj
Samanta and Supriyo Datta for many useful discussions.

\vspace{0.1in}

$^1$ Present address: Intel Corporation, RN2-40, 2200 Mission College
Blvd.  Santa Clara, CA 95052.

$^2$ Present address: Xilinx, 2100 Logic Dr., San Jose, CA 95124.

$^3$ Present Address: Yale University, Department of Electrical
Engineering, New Haven, CT 06520.

$^4$ Present address: Dept. of Physics, University of North Florida,
Jacksonville, FL 32224.

\end{document}